\documentclass[twocolumn,trackchanges]{aastex631}
\usepackage{float}
\usepackage{placeins}
\usepackage{graphicx}

\usepackage{comment}
\usepackage[subdued,defaultmathsizes]{mathastext}
\usepackage{natbib}  
\usepackage{amsmath}
\usepackage{amsfonts}
\usepackage{amssymb}
\usepackage{array}
\newcommand{\RNum}[1]{\uppercase\expandafter{\romannumeral #1\relax}}
\Mathastext [typewriter]

\begin{document}

\title{ExoGemS The Effect of Offsets from True Orbital Parameters\\ on Exoplanet High-Resolution Transmission Spectra}

\author[0009-0002-4295-9239]{Yasmine J.~Meziani}
\affiliation{Division of Physics, Mathematics, and Astronomy, California Institute of Technology, Pasadena, CA 91125, USA}
\affiliation{Department of Astronomy \& Carl Sagan Institute, Cornell University, Ithaca, NY 14853, USA}

\author[0000-0001-6362-0571]{Laura Flagg}
\affiliation{Department of Astronomy \& Carl Sagan Institute, Cornell University, Ithaca, NY 14853, USA}
\affiliation{Department of Physics and Astronomy, Johns Hopkins University, 3400 N. Charles Street, Baltimore, MD 21218, USA}

\author[0000-0001-7836-1787]{Jake D.~Turner}
\affiliation{Department of Astronomy \& Carl Sagan Institute, Cornell University, Ithaca, NY 14853, USA}

\author[0000-0001-9796-2158]{Emily K.~Deibert}
\affiliation{Gemini Observatory/NSF’s NOIRLab, Casilla 603, La Serena, Chile}

\author[0000-0001-5349-6853]{Ray Jayawardhana}
\affiliation{Department of Physics and Astronomy, Johns Hopkins University, 3400 N. Charles Street, Baltimore, MD 21218, USA}

\author[0000-0002-4451-1705]{Adam B. Langeveld}
\affiliation{Department of Astronomy \& Carl Sagan Institute, Cornell University, Ithaca, NY 14853, USA}
\affiliation{Department of Physics and Astronomy, Johns Hopkins University, 3400 N. Charles Street, Baltimore, MD 21218, USA}

\author[0000-0001-6391-9266]{Ernst J. W. de Mooij}
\affiliation{Astrophysics Research Centre, Queen's University Belfast, Belfast BT7 1NN, UK}



\begin{abstract}

High-resolution spectroscopy (HRS) plays a crucial role in characterizing exoplanet atmospheres, revealing detailed information about their chemical composition, temperatures, and dynamics. However, inaccuracies in orbital parameters can affect the result of HRS analyses. In this paper, we simulated HRS observations of an exoplanet's transit to model the effects of an offset in transit midpoint or eccentricity on the resulting spectra. We derived analytical equations to relate an offset in transit midpoint or eccentricity to shifted velocities, and compared it with velocities measured from simulated HRS observations. Additionally, we compared velocity shifts in the spectrum of the ultra-hot Jupiter WASP-76b using previously reported and newly measured transit times. We found that transit midpoint offsets on the order of minutes, combined with eccentricity offsets of approximately $0.1$, lead to significant shifts in velocities, yielding measurements on the order of several kilometers per second. Thus, such uncertainties could conflate derived wind measurements. 

\end{abstract}



\section{Introduction} \label{sec:intro}

The spectrum of an exoplanet reveals the physical, chemical, and biological processes that have shaped its history and govern its future. High-resolution spectroscopy (HRS) helps to extract and isolate the exoplanet's spectrum. It also simultaneously characterizes the planet's atmosphere due to its sensitivity to the depth, shape, and position of the planet's spectral lines \citep[e.g.,][]{snellen2010orbital, brogi2012signature, brogi2014carbon, brogi2016rotation, rodler2013detection, de2013detection, birkby2013detection, birkby2017discovery, birkby2018exoplanet, wyttenbach2015spectrally, schwarz2016slow, nugroho2017high, nugroho2020detection, hawker2018evidence, hoeijmakers2018atomic, hoeijmakers2019spectral, cauley2019atmospheric, wardenier2021decomposing}, as well as the temperature-pressure profile \citep{birkby2018exoplanet, brogi2019retrieving, ridden2023high, borsato2024small}. 

Since its initial use for exoplanet atmospheres in 2008 \citep{redfield2008sodium, snellen2008ground}, HRS has allowed for the important discoveries of various molecules in exoplanet atmospheres such as water \citep[e.g.,][]{birkby2013detection, birkby2017discovery, wehrhahn2023high}, sodium \citep[e.g.,][]{wyttenbach2015spectrally}, and carbon monoxide \citep[e.g.,][]{wehrhahn2023high}. Previously, the isolation of an exoplanet's atmospheric composition was achieved using low-resolution spectroscopy (LRS) \citep[e.g.,][]{ khalafinejad2021probing, genest2022effect, bocchieri2023detecting}. 

While LRS allows us to target major sources of opacity for exoplanet atmospheres (primarily H${}_2$O, but also CH${}_4$, HCN, and NH${}_3$), it faces ambiguities when multiple species overlap, making it challenging to identify molecules and determine their abundances. This is where the use of HRS becomes essential \citep{brogi2017framework}.

HRS and LRS differ significantly when examining velocity shifts in exoplanet atmospheres. LRS, with its broader spectral features, can only detect Doppler shifts larger than the typical velocities of exoplanets, providing a less detailed overview of atmospheric motions \citep{birkby2018exoplanet}. In contrast, HRS, with its high spectral resolution, is capable of detecting fine Doppler shifts, allowing for precise measurements of atmospheric dynamics, such as wind patterns and rotational velocities \citep{birkby2018exoplanet}. This allows for detailed mapping of day-night winds and atmospheric circulation, which are crucial for understanding the atmospheric behavior and climate of exoplanets.

However, high-resolution spectra require an assumed set of orbital parameters to extract an exoplanet's spectrum. Orbital parameters, like any other measured quantities, have uncertainties that can cause deviations from their true values. This is particularly significant for HRS when there is an offset in transit midpoint or eccentricity. An example can be found in \citet{asnodkar2022variable}, where after using HRS to measure day-to-nightside winds in the atmosphere of KELT-9b, the researchers discovered that utilizing an updated ephemeris led to different wind measurements. 

Day-night winds, are claimed due to a systemic velocity blueshift,\citep[e.g.,][]{ snellen2010orbital, kempton2012constraining, brogi2016rotation, kesseli2021confirmation, wardenier2021decomposing, asnodkar2022variable}. However, it is also plausible that inaccuracies in ephemeris and eccentricity measurements could contribute to a false detection of day-night winds on an exoplanet.

In this paper, we will use both real and simulated data to examine how offsets in transit midpoint or eccentricity affect the resulting analysis of the exoplanet's atmosphere. In Section \ref{sec: data}, we will describe the simulation of our high-resolution optical spectra as well as additionally describing our high-resolution optical spectra for WASP-76b obtained using GRACES. Data analysis methods, including analytical derivations and numerical methods, are presented in Section \ref{sec: methods}. The results and discussion are described in Sections \ref{sec: results} and \ref{sec: discussion}, with a conclusion in Section \ref{sec: conclusion}.

\section{Data} \label{sec: data}
We created simulated observations using real GRACES spectra of a transit of WASP-85Ab. GRACES \citep{chene2021dragraces} uses a fiber optic feed to combine the large collecting area of the Gemini North Telescope with the high resolving power of the ESPaDOnS (Echelle SpectroPolarimetric Device for the Observation of Stars; \citet{donati2003espadons}) spectrograph at the Canada France Hawaii Telescope (CFHT). GRACES achieves a maximum resolution power of $R \sim 60,000$ and provides wavelength coverage from $400\;\mathrm{nm}$ to $1000\;\mathrm{nm}$ \citep{chene2021dragraces}.

We reduced the spectra using OPERA, the Open Source Pipeline for ESPaDOnS Reduction and Analysis \citep{martioli2012open}. We then insert CrH absorption signals into each in-transit observation \footnote{\label{foot: Laura's code}\url{https://github.com/lauraflagg/svd_exoplanets}}. We create CrH transmission spectra templates using the methods in \citet{flagg2023exogems}, which involve the TRIDENT radiative transfer code \citep{macdonald2022trident} and the CrH line list from \citet{burrows2002new}. We use a CrH template due to its availability and because the molecule is absent in WASP-85Ab (Flagg et al., in prep). While CrH is used as an example, there is no wavelength dependence in any of the equations, meaning that our discoveries and methods are broadly applicable to any molecule, atom, or species.

We then processed the data as it would be for other high-resolution exoplanet transmission spectra. We remove the telluric and stellar features with Singular Value Decomposition (SVD) and then shift the spectra to the stellar rest frame based on the barycentric correction calculated with \texttt{astropy.time}.

While we relied on simulated signals for WASP-85Ab, we also analyzed real observational data from a different target, WASP-76b. We observed one transit of WASP-76b with GRACES \citep{chene2021dragraces}. The data of WASP-76b obtained with GRACES was published in \citet{deibert2021detection, deibert2023exogems}. \citet{deibert2021detection} recovered absorption features due to neutral sodium and reported a new detection of the ionized calcium triplet at $\sim 850\: \mathrm{nm}$ in the atmosphere of WASP-76b. 

\section{Methods} \label{sec: methods}
\subsection{CCF matrix} \label{subsec: CCF matrix}
\subsubsection{CCF matrix creation and fitting the signal}
\label{subsubsec: creation and fitting}

To evaluate how our simulated data responds to an offset in transit midpoint or eccentricity, we generate a CCF matrix for different offsets in ephemerides or eccentricities. We use standard methods for creating the CCF matrix \citep[e.g.,][]{birkby2013detection, rodler2014feasibility, flagg2023exogems}. To summarize, we shift the spectra to the planetary rest frame using

\begin{equation}\label{eq:vshift_eph}
v_{shift} = K_p \sin\left( \frac{ 2 \pi }{P} \left(t-t_0\right)\right).
\end{equation}

We chose $K_{p}$ as a free parameter, allowing it to vary between $0$ and $250~\mathrm{km\: s^{-1}}$. Negative values for $K_{p}$ are physically unrealistic, and because the data were simulated, we selected this range to ensure that the signal could be clearly observed. Once shifted, we cross-correlate the spectra with the template before coadding the CCFs. This procedure was implemented using a custom code\textsuperscript{\ref{foot: Laura's code}}. 

Next, we fit a two-dimensional Gaussian to the CCF matrix using the \(\mathtt{lmfit}\)\footnote{\label{foot:lmfit}\url{https://lmfit.github.io/lmfit-py/}} Python package. The x coordinate of the center of the Gaussian, the y coordinate of the center of the Gaussian, the amplitude, the width of the Gaussian along the x-axis, the width of the Gaussian along the y-axis, and the Gaussian rotation angle are set as free parameters. The measured Doppler shift and its uncertainty are the best-fit x-center. 

\subsubsection{Offset in transit midpoint} \label{subsubsec: methods_ephoffset}
We determined \(v_{sys}\) (systemic velocity) and \(K_p\) (planetary radial velocity semi-amplitude), for ephemeris offsets spanning from -16 minutes to 16 minutes with intervals of 2 minutes.
We chose this range to ensure we had enough data points to accurately plot the correlation between measured systemic velocities and ephemeris offset. We formulated an analytical expression that establishes a relationship between measured velocity shifts and ephemeris offsets, facilitating predictions regarding the impact of ephemeris offsets on CCF matrix velocity shifts. 

To derive our analytical approximation, we first shift all spectra into the planetary rest frame using Equation \ref{eq:vshift_eph}, where \(P\) is equal to the planet's period, $t$ is the date, \(t_0\) is the true midpoint date, and $v_{shift}$ is the velocity shift. Next, we subtract the \(v_{shift}\), accounting for the offset in the midpoint \(dt_{0}\) from the initially shifted \(K_p\).

\begin{equation}\label{eq:dvshift}
\begin{split}
\Delta v_{shift,eph} = K_p \sin( \frac{2 \pi }{P} (t-t_0+dt_{0}))\\
-K_p\sin( \frac{ 2 \pi }{P} (t-t_0)).
\end{split}
\end{equation}

We assume a small angle approximation for both sine functions which results in the equation 

\begin{equation}\label{eq:post small angle}
\Delta v_{shift,eph} =K_{p} \cdot \frac{2\pi}{P} (t-t_{0}+dt_{0}) - K_{p} \cdot \frac{2\pi}{P} (t-t_{0}).
\end{equation}

After simplifying Equation \ref{eq:post small angle}, we obtain our final analytical solution

\begin{equation}\label{eq:dvshift_final}
\Delta v_{shift, eph} = \frac{K_p 2 \pi }{P} dt_{0}.
\end{equation}

\subsubsection{Offset in eccentricity}\label{subsubsec: methods_eccoffset}

We analyzed simulated data of WASP-85Ab to explore the relationship between the CCF matrix and velocity shifts for different offsets in eccentricity. Using techniques described in the CCF matrix section \ref{subsec: CCF matrix}, we measured $v_{sys}$ and $K_{p}$ for a range of eccentricity offsets from $-0.05$ to $0.08$, and $-0.12$ to $0.08$, with intervals of $0.01$. We set the lower bound of the eccentricity offset at $-0.05$ and $-0.12$ because the simulated data had a true eccentricity of $0.05$ and $0.12$, respectively. Since eccentricity cannot be negative, a bound lower than $-0.05$ or $-0.12$ is not possible. We selected an upper bound of $0.08$ to ensure a sufficient number of data points for plotting the correlation between measured systemic velocities and eccentricity offset. Similarly, as mentioned in \ref{subsubsec: methods_ephoffset}, we developed an analytical expression to correlate measured velocity shifts with eccentricity offsets.

A closed-form analytical solution does not exist because Kepler's equation is transcendental, involving a non-algebraic relationship between the mean anomaly $M$, which is defined as $\frac{2\pi}{P(t-t_{0})}$, and the eccentric anomaly $E$ \citep{murray1999solar}; instead, we used a numerical approach. The derivation below uses methods outlined in \citet{kohout1972optimized}, \citet{montalto2011exoplanet}, and \citet{grant2020uncovering}. We first begin with Kepler's equation

\begin{equation}\label{eq:mean_anomaly}
M = E - e\sin E,
\end{equation}

where $e$ is the eccentricity. We solve for $E$ using the Newton-Raphson method. Using our value for $E$, we solve for the true anomaly
\begin{equation}\label{eq:true_anomaly}
\nu(e)= 2 \cdot \arctan\left(\sqrt{-\frac{e + 1}{e - 1}} \cdot \tan\left(\frac{E(e)}{2}\right)\right).
\end{equation}
From Equation \ref{eq:true_anomaly}, we can solve for the radial velocity (RV) as a function of eccentricity in the planet's rest frame,

\begin{equation}\label{eq:radial_velocity}
v_{shift,ecc}(e)= K_{p}[\cos(\omega+\nu(e))+e \cdot \cos(\omega)],
\end{equation}

where $K_{p}$ is the semi-amplitude of the orbiting planet and $\omega$ is the angle of periastron. 
\\~\\
To estimate how a slight offset in eccentricity affects the atmospheric signal, we subtract Equation \ref{eq:radial_velocity} from the shifted radial velocity, adding the offset in eccentricity ($v_{shift}(e+de)$)

\begin{equation}\label{eq: vshift_ecc}
\begin{split}
\Delta v_{shift,ecc} =  K_{p}[\cos(\omega+\nu(e+de))\\
-\cos(\omega+\nu(e))+de\cdot \cos(\omega)].
\end{split}
\end{equation}

Note that $K_{p}$ is a function of $e$ as well \citep{basilicata2024gaps}; however, for the eccentricity offsets considered here, the resulting error is on the order of $1\%$. 

\subsection{WASP-76b Observations}\label{subsec:methods_WASP-76}

\begin{table*}[t]
\centering
\begin{tabular}{|l|l|l|l|}
\hline
\textbf{Parameter} & \textbf{Units} & \textbf{Value} & \textbf{Source} \\
\hline
Period & days & $1.80988198$ & \cite{ehrenreich2020nightside} \\
Period & days & $1.809880580$ & \cite{ivshina2022tess} \\
Mid-transit time & BJD & $2458080.626165$ & \cite{ehrenreich2020nightside} \\
Mid-transit time & BJD & $2458080.6257$ & \cite{ivshina2022tess} \\
\hline
\end{tabular}
\caption{Orbital period and ephemeris measurements for WASP-76b between the earlier data reported by \citet{ehrenreich2020nightside} and the updated values from \citet{ivshina2022tess}. We do not consider uncertainties because they are not typically used when calculating transmission spectra.}
\label{table:1}
\end{table*}

We generated two transmission spectra: one using the older ephemeris from \citet{ehrenreich2020nightside}, as employed by \citet{deibert2021detection}, and the other using the updated ephemeris from \citet{ivshina2022tess}. After reducing the data as described in Section \ref{sec: data}, we followed the methods outlined in \citet{wyttenbach2015spectrally} and \citet{turner2020detection} to extract the planetary signal in the form of a transmission spectrum in the planet's rest frame. First, we reduced the data observed with GRACES as described in \ref{sec: data}. We then shifted the spectra to the stellar rest frame using systemic and stellar radial velocity values from \cite{west2016three} and \cite{ehrenreich2020nightside}, following the approach of \citet{deibert2021detection}. Next, we applied the radial velocity semi-amplitude values for the planet, either from \citet{ehrenreich2020nightside} or derived directly from \citet{deibert2021detection} (for each detected species), to shift the spectra to the planet's rest frame. Once the spectra were in the planet’s rest frame, we summed the in-transit frames to create the transmission spectrum. Further details of the process can be found in \citet{deibert2021detection}.

We fitted one-dimensional Gaussians using \(\mathtt{lmfit}\) to the transmission spectra for both ephemerides to find the centers of the absorption features.
We compared the spectra for Ca II and Na I, and their resulting line centers, using an old ephemeris from \citet{ehrenreich2020nightside}, and a newer one from \citet{ivshina2022tess}.

\section{Results} \label{sec: results}

\subsection{Simulated Data} \label{subsec: results simulated}

\subsubsection{Offset in transit midpoint} \label{subsubsec: results ephemeris}
\begin{figure}[h] 
  \centering
  \includegraphics[width=1.0\linewidth]{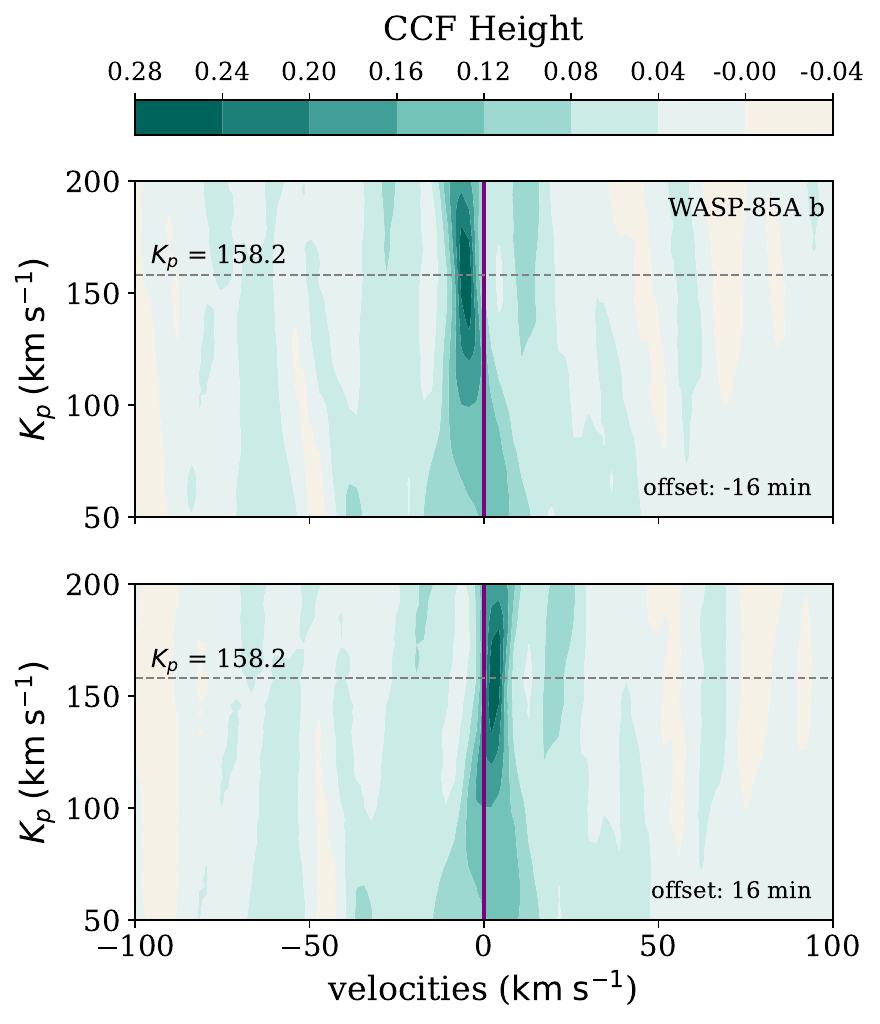}
  \caption{The results of cross-correlating the transmission spectrum—generated after injecting a CrH spectral signature into the data—with a model of WASP-85Ab's spectral signal using different offsets in transit midpoint are shown. The top plot displays the 2D $K_{p}$-$v_{sys}$ map (also known as the CCF matrix) for an ephemeris offset of -16 minutes, while the bottom plot shows the same for an offset of 16 minutes. We observe that larger transit midpoint offsets correspond to higher systemic velocities, indicating a positive correlation.}
  \label{fig:ephemeris}
\end{figure}

\label{subsec: results_eph_off}
\begin{figure}[h]
    \centering
    \includegraphics[width=1.0\linewidth]{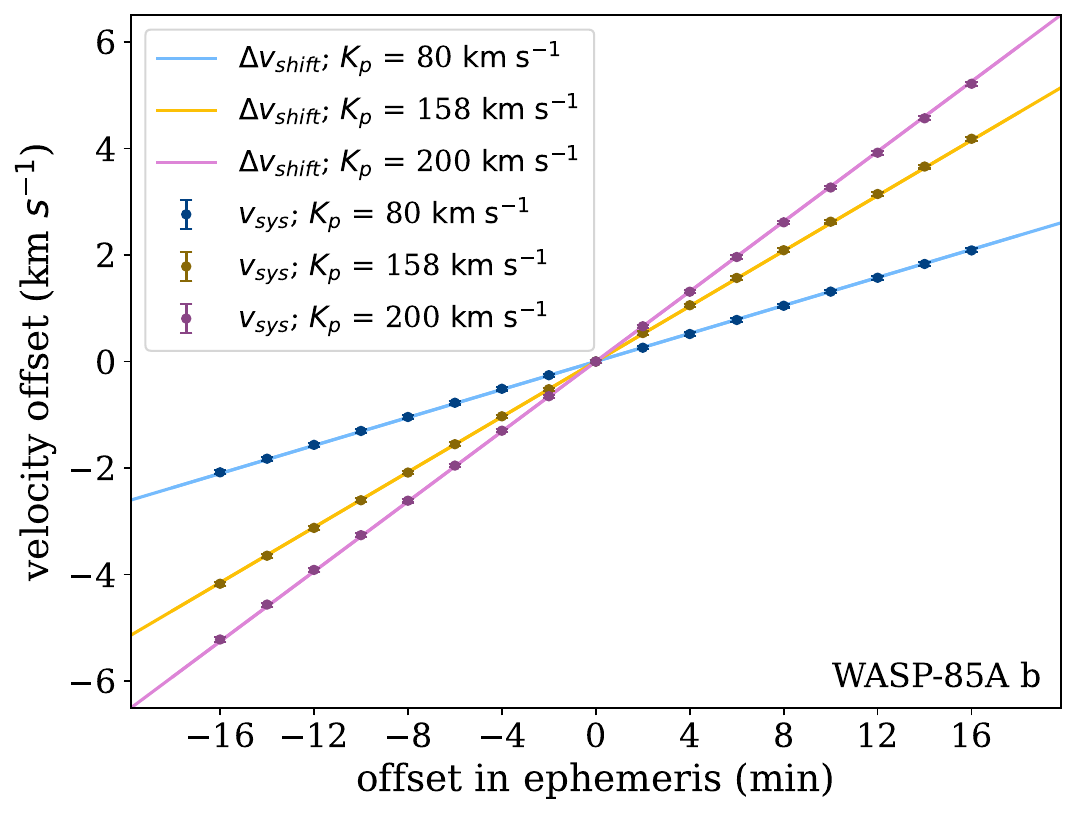} 
    \caption{Our analytical solution from Equation~\ref{eq:dvshift_final}, along with our measured systemic velocities---obtained by calculating the centers of our simulated observations using the methods outlined in Section~\ref{subsec: CCF matrix} for $K_{p}$ values of 80, 158, and 200---are shown. The straight lines represent our analytical solution using Equation~\ref{eq:dvshift_final}, while the dotted points represent the measured systemic velocities obtained from our simulated observations. The 1$\sigma$ error bars indicate the uncertainty in velocity measurements, which were automatically calculated by the code\textsuperscript{\ref{foot: Laura's code}} used to determine the centers of the cross-correlation function (CCF) matrices. Small offsets in the transit midpoint, on the scale of minutes, result in significant velocity shifts of several $\mathrm{km\: s^{-1}}$. Moreover, increasing the $K_{p}$ values amplifies these velocity offsets.}
    \label{fig:lineplots}
\end{figure}

As described in \ref{subsubsec: creation and fitting}, we generate a cross-correlation function (CCF) matrix for different offsets in ephemerides. In Figure \ref{fig:ephemeris} we plot the CCF matrix for two different offsets, one at $16$ minutes before and one at $16$ minutes after the true transit midpoint. These plots show an increasing shift in systemic velocities when increasing the offset in ephemeris. In Figure \ref{fig:lineplots}, we plot the analytical solution described in Section \ref{sec: methods}, along with the velocity shifts measured from the CCF matrices generated from the simulated data. As shown in Figure \ref{fig:lineplots}, our analytical solution aligns almost perfectly with the systemic velocities measured from our simulated data. 

From Figure \ref{fig:lineplots}, we see that small ephemeris offsets, on the order of minutes can lead to velocity shifts of several kilometers per second. Note that the relationship between ephemeris offsets and velocity shifts is linear and proportional to the $K_{p}$, which matches the analytical solution in Equation \ref{eq:dvshift_final}. 

\begin{figure*}[t]
  \centering
  \includegraphics[width=1.0\linewidth]{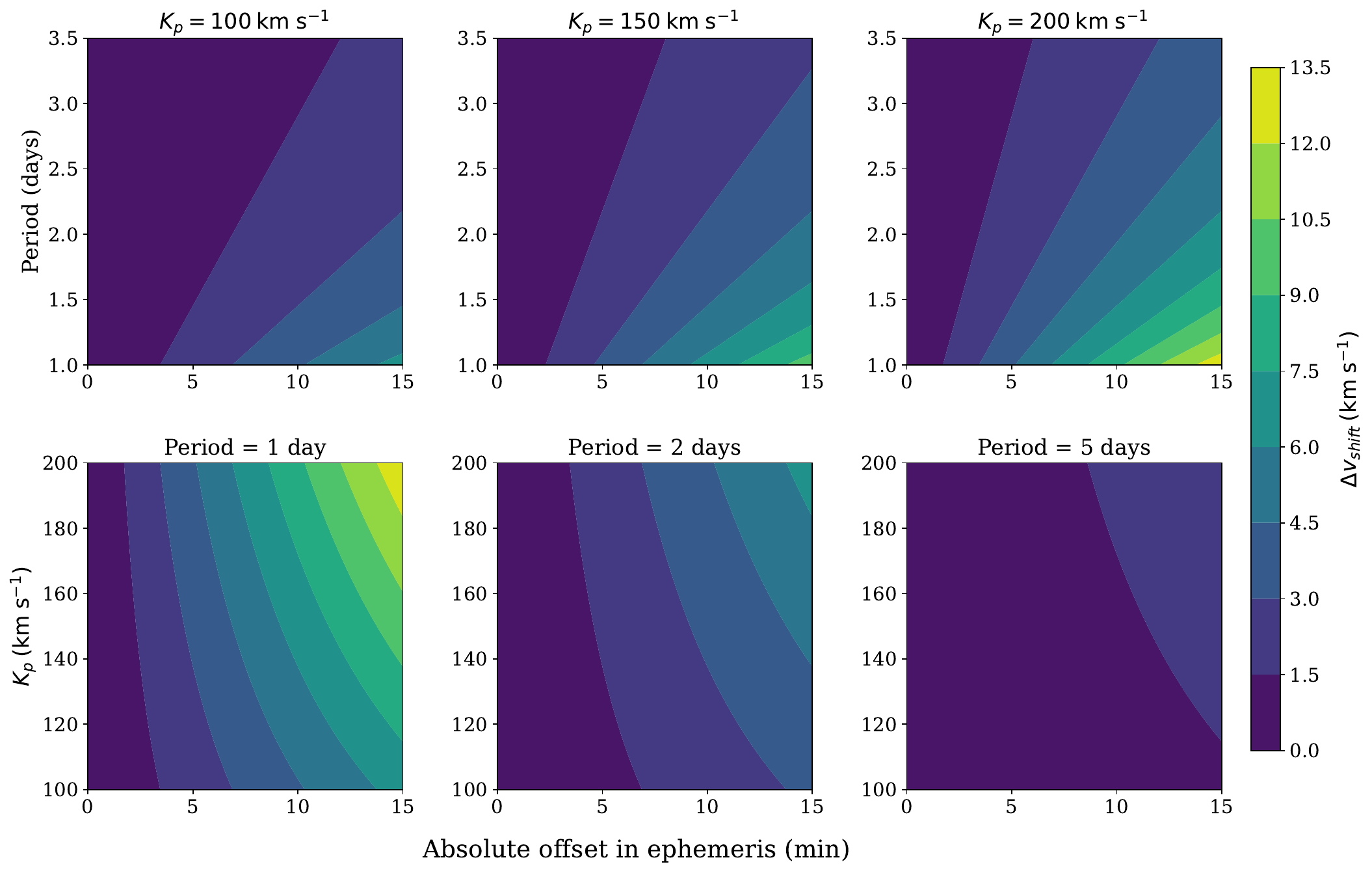}
  \caption{Contour plots to approximate $v_{shift}$ for different offsets in ephemeris. The top three panels of the figure use fixed $K_{p}$ values of $100 \;\mathrm{km\: s^{-1}}$, $150 \;\mathrm{km\: s^{-1}}$, and $200 \;\mathrm{km\: s^{-1}}$, while varying the period and ephemeris offset. The bottom three panels use fixed period values of $1 \; \mathrm{day}$, $2 \; \mathrm{days}$, and $5 \; \mathrm{days}$, while varying the $K_{p}$ and ephemeris offset.}
  \label{fig:unitconversions}
\end{figure*}

Our final analytical solution can also be represented as

\begin{equation}\label{eq:simplified}
    \begin{split}
        \Delta v_{shift,eph} \approx 4.36\cdot10^{-3}\,[K_{p}\,(\mathrm{km\: s^{-1}})
        \cdot \frac{1}{P\,(\mathrm{day})} \\
        \cdot dt_{0}\,(\mathrm{min})]\,\;\mathrm{km\: s^{-1}},
    \end{split}
\end{equation}

where $4.36\cdot10^{-3}$ is a constant accounting for unit conversions. In Figure \ref{fig:unitconversions}, we show contour plots that estimate the velocity offset for various combinations of orbital period and $K_{p}$ values. 


\subsubsection{Offset in eccentricity}\label{subsubsec: results eccentricity}




\begin{figure}
    \centering

    \includegraphics[width=0.95\columnwidth]{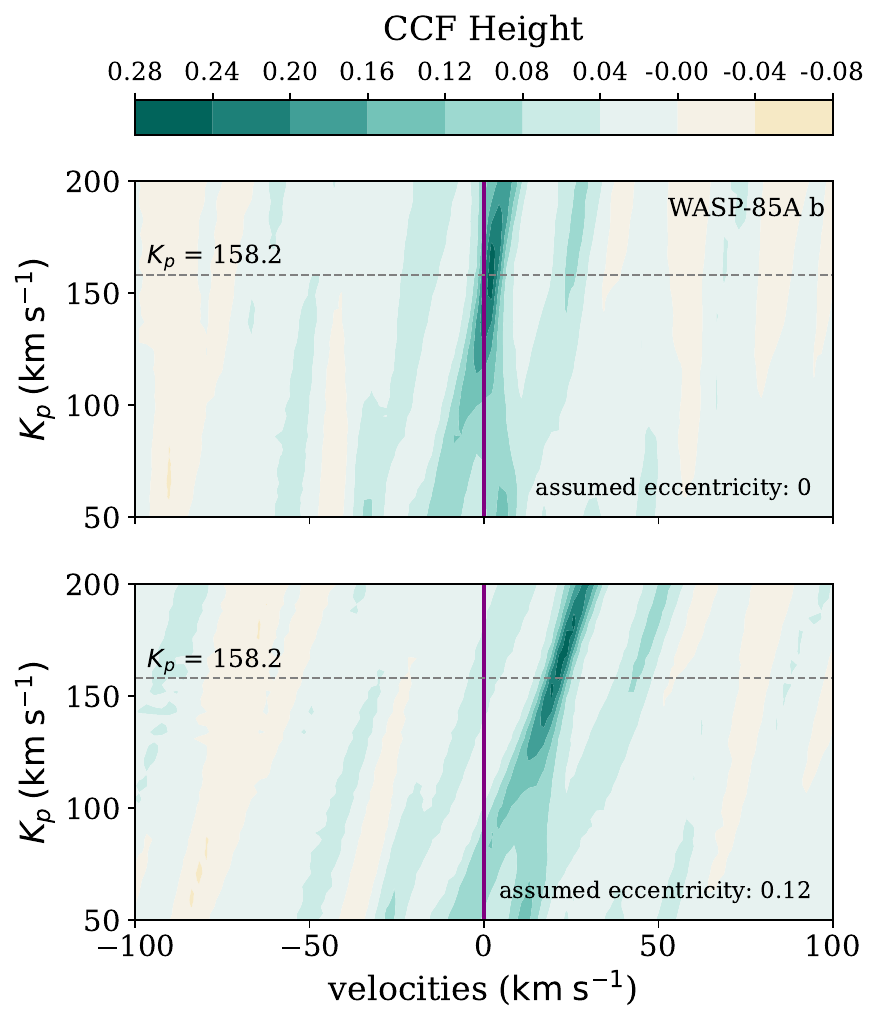}
    \raggedright
    \par
    (a) We cross-correlated the transmission spectrum---generated by injecting a CrH spectral signature into the data---with a model of WASP-85Ab's spectral signal, using eccentricity offsets of $-0.05$ and $0.08$ relative to the true value of $0.05$.
    \par
    \label{subfig:ecc_0.05}

    \vspace{0.5cm} 

    \includegraphics[width=0.95\columnwidth]{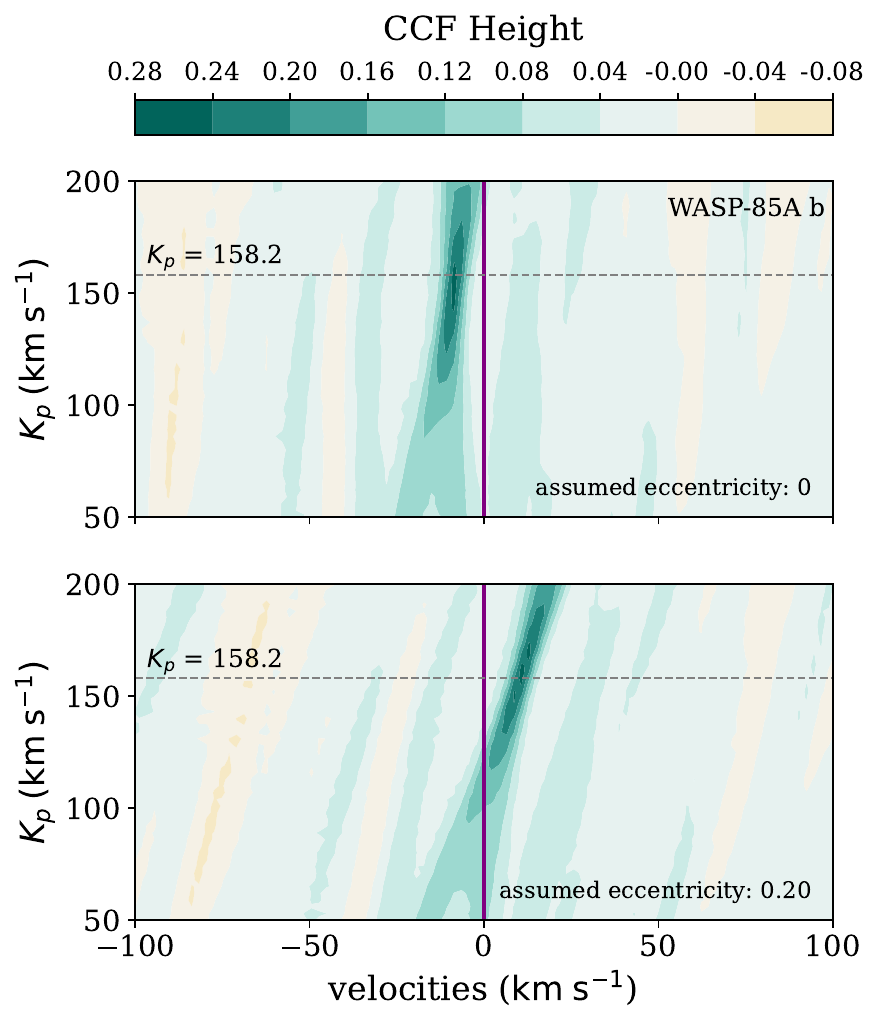}
    \raggedright
    \par
    (b) The same as Figure \ref{fig:ecc_offsets}a, but for a true eccentricity of $0.12$ with offsets of $-0.12$ and $0.08$.
    \par
    \label{subfig:ecc_0.12}

    \caption{Cross-correlation results showing the effects of eccentricity offsets.}
    \label{fig:ecc_offsets}
\end{figure}

\begin{figure*}[t!] 
  \centering
  \includegraphics[width=1.0\linewidth]{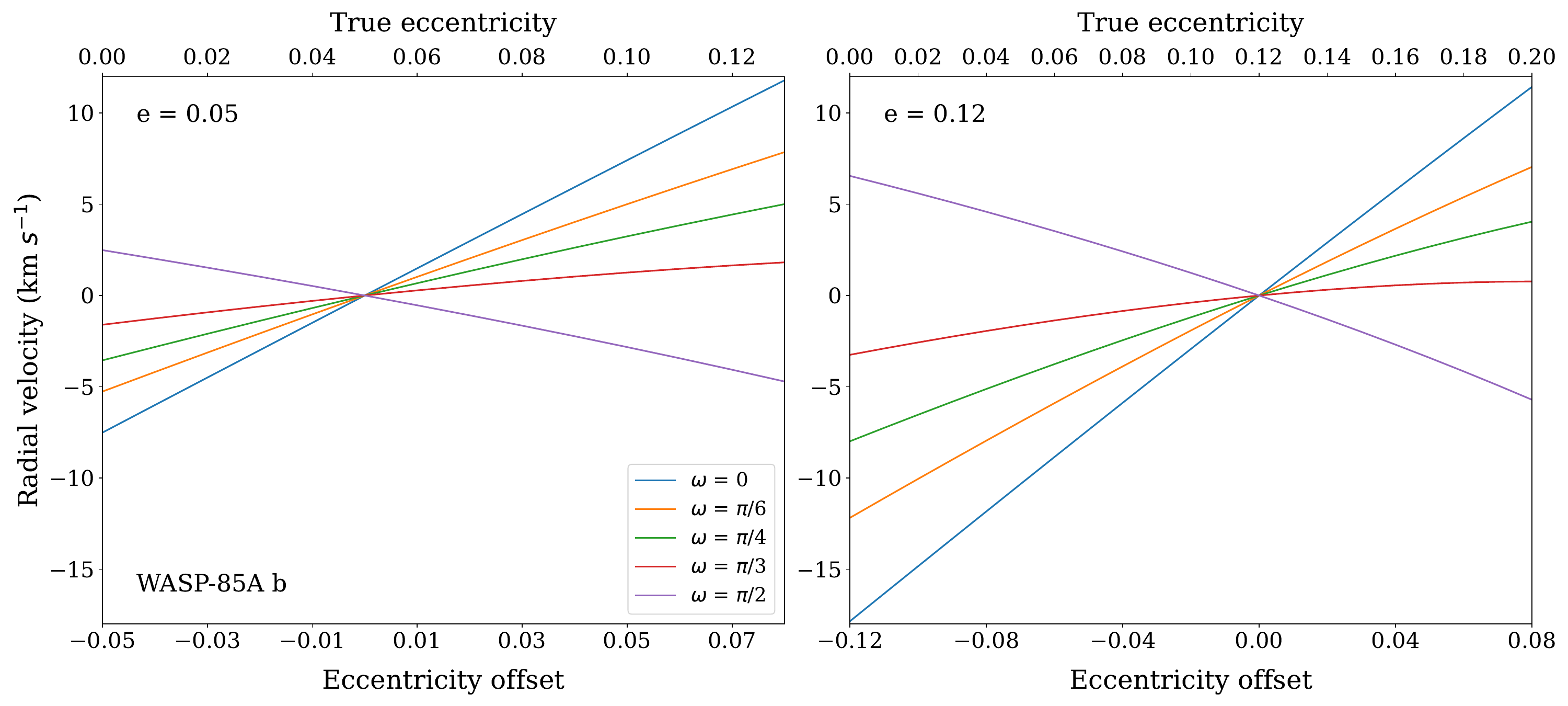}
  \caption{The planetary radial velocity shifts for different eccentricity offsets for several values of the angle of periastron ($\omega$). The panel on the right plots the radial velocity for different offsets in eccentricity for an original eccentricity value $e = 0.12$, while the panel on the left does the same for an original eccentricity value of $e = 0.05$. We can see that small offsets in eccentricity on the order of $0.01$ lead to offsets in radial velocities on the order of several $\mathrm{km\: s^{-1}}$}.
  \label{fig:ecc line plots}
\end{figure*}

\begin{figure}
  \centering
  \includegraphics[width=1.0\linewidth]{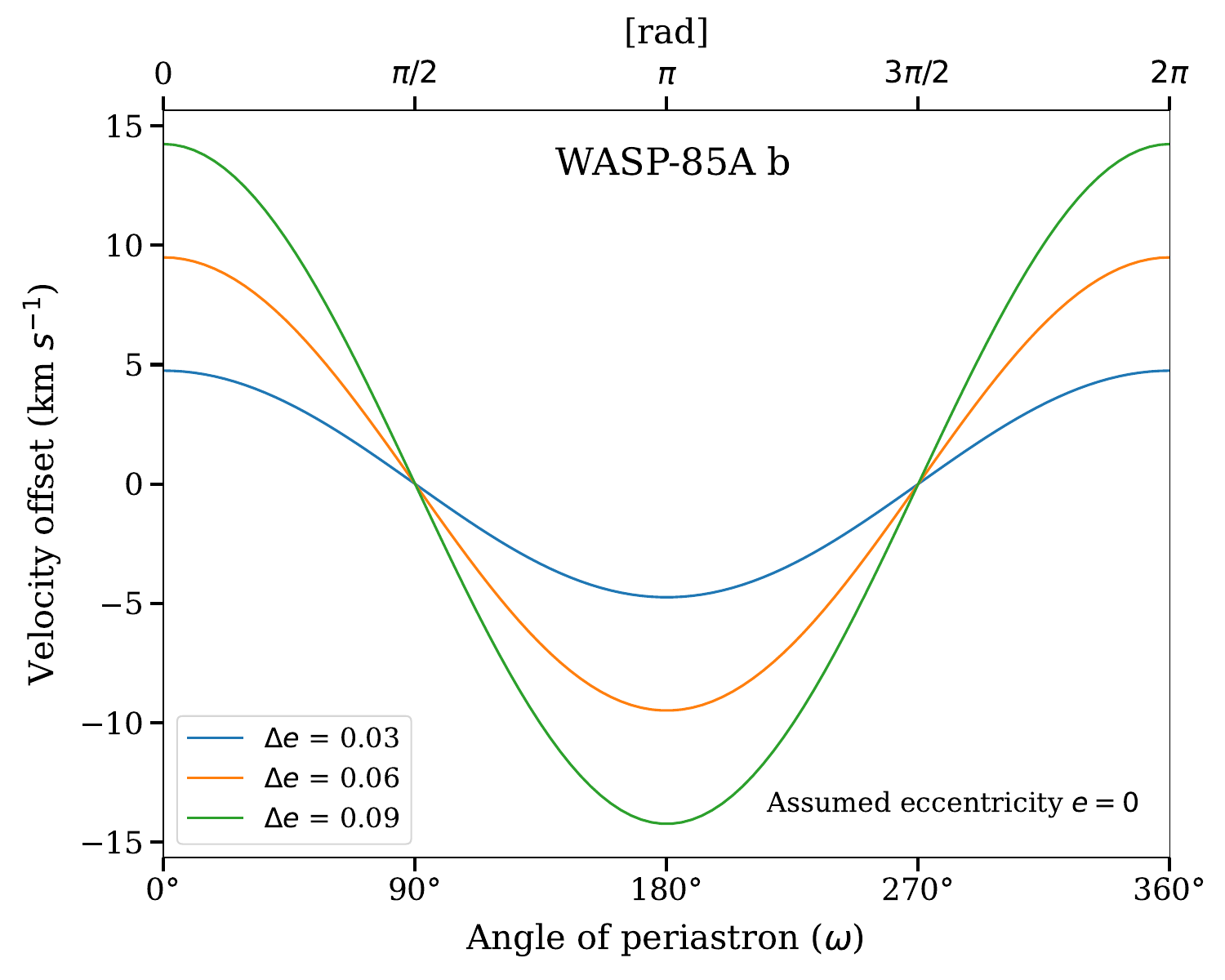}
  \caption{The figure above illustrates the relationship between velocity offset and angle of periastron for an assumed eccentricity of zero ($e=0$). Each line represents the relationship between velocity offset and angle of periastron for different true eccentricity offset. Angles of periastron at $\frac{\pi}{2}$ and $\frac{3 \pi}{2}$ correspond to zero velocity offset, consistent with exoplanet conventions.}
  \label{fig: angle periastron}
\end{figure}

Following the methodology in Subsection \ref{subsubsec: methods_eccoffset}, we introduced eccentricity offsets into the simulated HRS observations of WASP-85A~b for true orbital eccentricities of $e = 0.05$ and $e = 0.12$. For the dataset with $e = 0.05$, we varied the offset from $-0.05$ to $0.08$ in steps of $0.01$, generating a CCF matrix for each offset. The reasons for these bounds are explained in Subsection \ref{subsubsec: methods_eccoffset}. For the dataset with $e = 0.12$, we varied the offset from $-0.12$ to $0.08$ in steps of $0.01$, again generating a CCF matrix for each offset. We set the lower bound to $-0.12$ because eccentricity cannot be negative, and the upper bound to $0.08$ to ensure enough data points to analyze the correlation between measured systemic velocities and eccentricity offset.

In Figure \ref{fig:ecc_offsets} we plot a sample of these CCF matrices. Figure \ref{subfig:ecc_0.05} plots the CCF matrices using a true eccentricity of $e=0.05$ and Figure \ref{subfig:ecc_0.12} using a true eccentricity of $e=0.12$. We observe an increase in velocity as the eccentricity offset increases, depending on the angle of periastron. Consistent with \citet{montalto2011exoplanet}, who showed that an eccentricity of $e = 0.01$ can cause radial velocity shifts of several kilometers per second, our findings also indicate that eccentricities around $0.01$ significantly affect radial velocity measurements.

As seen in Equation \ref{eq: vshift_ecc}, the angle of periastron ($\omega$), also has an effect on the RV shift. Figure \ref{fig:ecc line plots} shows the relationship between eccentricity offset and radial velocity shift using Equation \ref{eq: vshift_ecc}, for the exoplanet WASP-85A b with true eccentricities of $0.05$ (Figure \ref{fig:ecc line plots}, left) and $0.12$ (Figure \ref{fig:ecc line plots}, right). Different color lines correspond to different angles of periastron, defined such that $\omega = \frac{3\pi}{2}$ places the periastron along the line-of-sight to the observer \citep{montalto2011exoplanet}. For WASP-85A b with a true eccentricity of $0.05$, an eccentricity offset between $-0.05$ and $0.08$ results in velocity shifts ranging from $-8$ to $12\;\mathrm{km\: s^{-1}}$, depending on the angle of periastron. Similarly, with a true eccentricity of $0.12$, an eccentricity offset between $-0.12$ and $0.08$ leads to velocity shifts between $-18$ and $12\;\mathrm{km\: s^{-1}}$, also depending on the angle of periastron.

Since it is common in exoplanet literature to assume an eccentricity of zero, Figure \ref{fig: angle periastron} shows the relationship between velocity offset and angle of periastron for an assumed eccentricity of zero with each line corresponding to how large the effect is for different true orbital eccentricities. It is evident (Figure \ref{fig: angle periastron}) that the angle of periastron affects how eccentricity offsets influence velocity offsets. 

\begin{figure*}[h]
  \centering
  \includegraphics[width=1.0\linewidth]{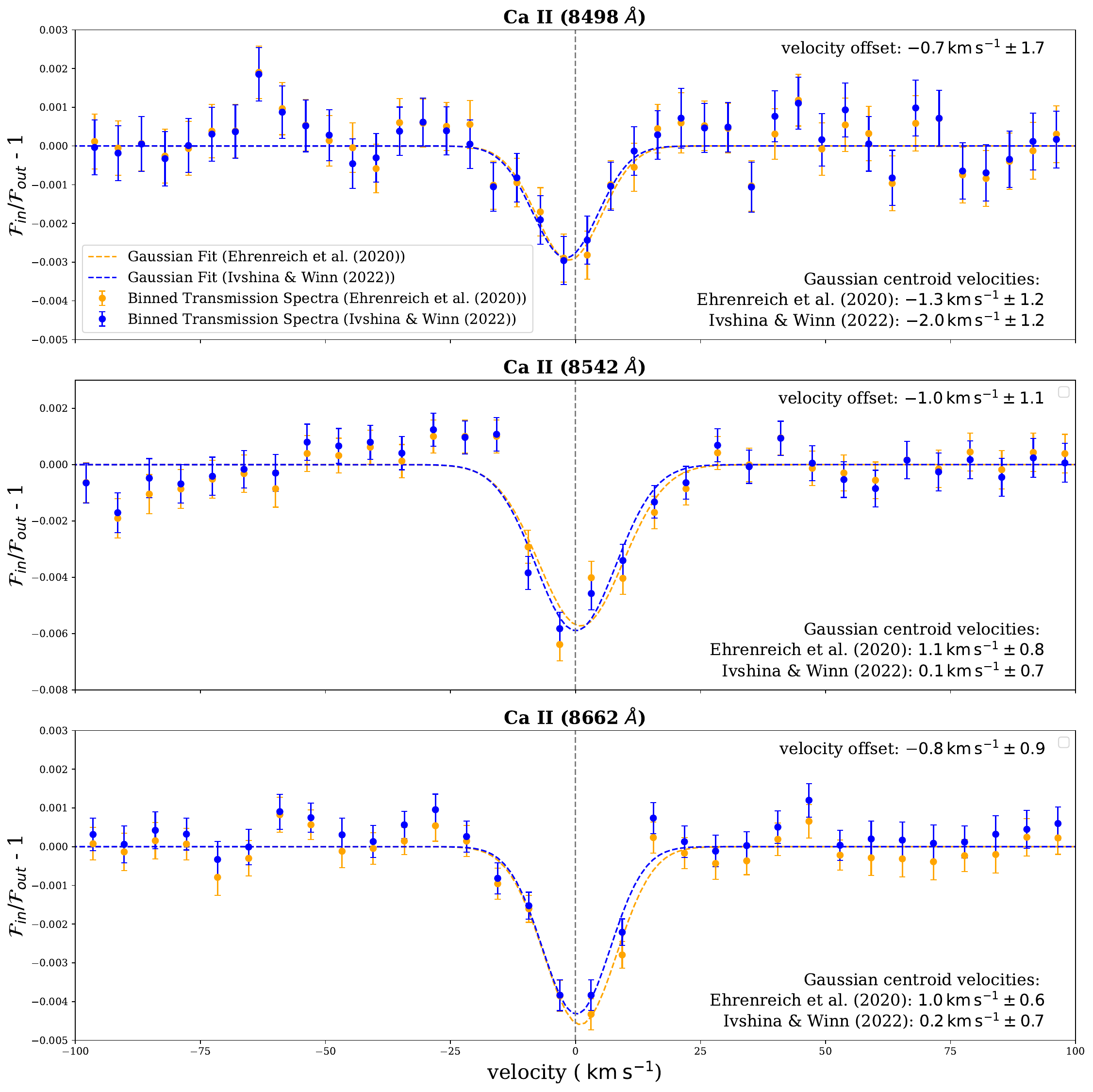}
  \caption{The top, middle, and bottom panels show GRACES transmission spectra of WASP-76b around the ionized calcium triplet at $849.8\:\mathrm{nm}$, $854.2\:\mathrm{nm}$, and $866.2\:\mathrm{nm}$, respectively. The points represent the binned transmission spectra, while the dotted lines show Gaussian fits to each line profile. Each panel displays results using the ephemeris from either \citet{ehrenreich2020nightside} or \citet{ivshina2022tess}. The velocity offset refers to the difference in the measured line centers resulting from the use of these two ephemerides.}
  \label{fig:calcium}
\end{figure*}

\subsection{WASP-76b} \label{subsec: results wasp}

\begin{figure*}[h]
  \centering
  \includegraphics[width=1.0\linewidth]{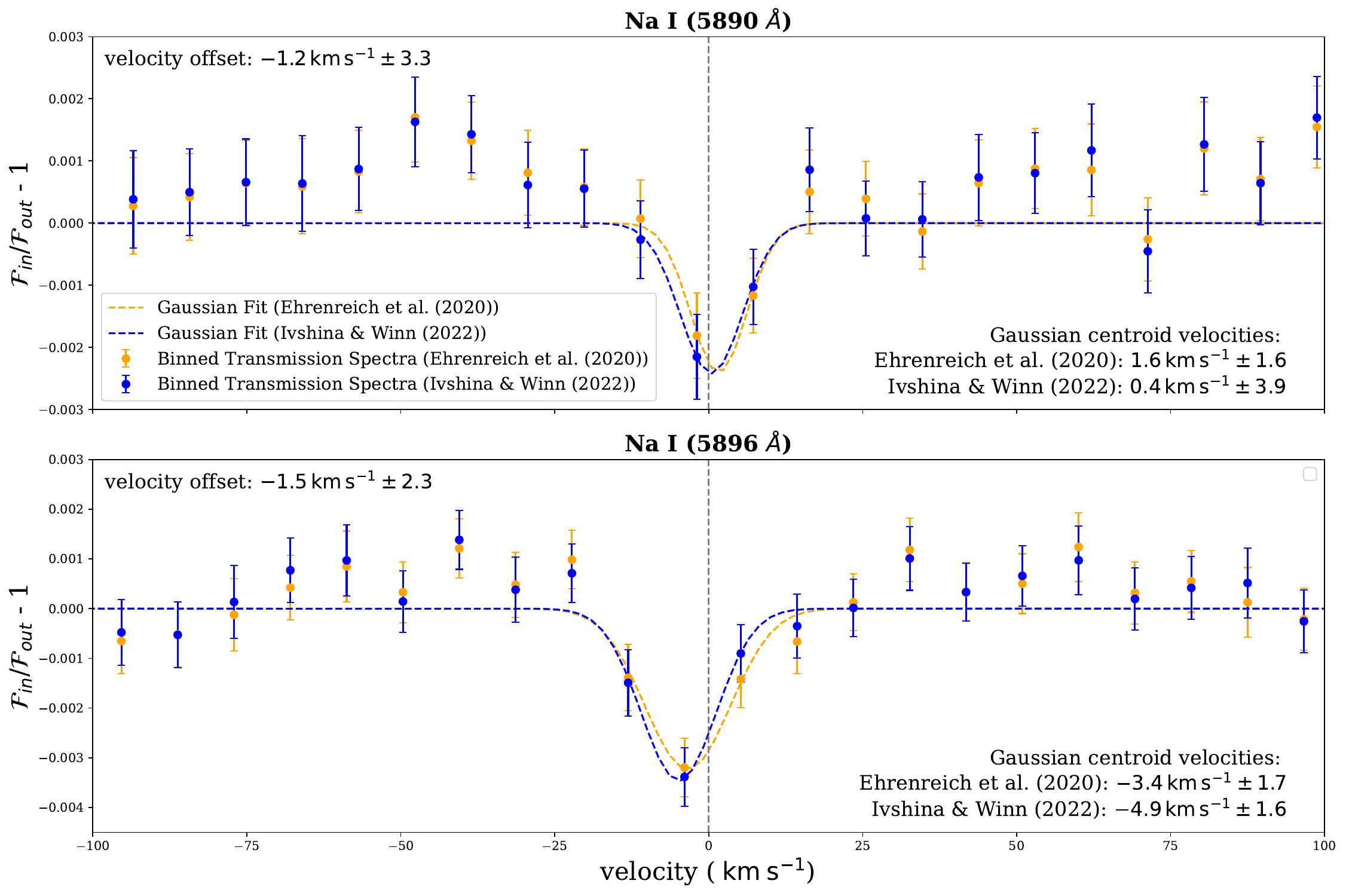}
  \caption{Similar to Figure \ref{fig:calcium}, but for the sodium doublet at $589.0$ and $589.6 \: \mathrm{nm}$. Using the updated ephemeris from \citet{ivshina2022tess}, we detect blueshifted features with velocity offsets of $-1.2 \pm 3.3 \;\mathrm{km\: s^{-1}}$ and $-1.5 \pm 2.3 \;\mathrm{km\: s^{-1}}$, respectively.}
  \label{fig:sodium}
\end{figure*}

In \citet{deibert2021detection}, the ionized calcium triplet and sodium doublet were detected using GRACES transmission spectra using the transit time from \citet{ehrenreich2020nightside}. When we applied the updated transit timing from \citet{ivshina2022tess}, velocity shifts were observed in all three spectral lines ($8498 ; \text{\AA}$, $8542 ; \text{\AA}$, and $8662 ; \text{\AA}$), caused by differences in the ephemeris solutions. Figure \ref{fig:calcium} shows the GRACES transmission spectra around the $8498 \; \text{\AA}$ calcium line, with a velocity shift of $\sim 0.7 \; \mathrm{km\: s^{-1}}$. Velocity shifts of $0.7$-$1 \;\mathrm{km\: s^{-1}}$ were observed across all three calcium lines, depending on the line (see Table \ref{tab:species_shifts}). In the Na I doublet, we also observed a velocity shift of $-1.2 \pm 4 \; \mathrm{km\: s^{-1}}$ in the D1 line and a shift of $-1.5 \pm 4 \; \mathrm{km\: s^{-1}}$ in the D2 line.

\begin{table}[ht] 
    \centering
    \begin{tabular}{|l|c|c|}
        \hline
        \textbf{Species} & \textbf{Wavelength (\AA)} & \textbf{Shift ($\; \mathrm{km\: s^{-1}}$)} \\
        \hline
        Ca II & 8498 & -0.7 $\pm$ 1.7\\
        \hline
        Ca II & 8542 & -1.0 $\pm$ 1.1\\
        \hline
        Ca II & 8662 & -0.8 $\pm$ 0.9\\
        \hline
        Na I & 5890 & -1.2 $\pm$ 3.3\\
        \hline
        Na I & 5896 & -1.5 $\pm$ 2.3\\
        \hline
    \end{tabular}
    \caption{The velocity shifts in the transmission spectra represent the differences in measured Gaussian fit centers when applying the ephemeris from \citet{ehrenreich2020nightside} versus that from \citet{ivshina2022tess}.}
    \label{tab:species_shifts}
\end{table}

To confirm whether this result correlates to our analytical equation (Equation \ref{eq:dvshift_final}), we first calculated the offset in transit midpoint timings between \citet{ehrenreich2020nightside} and \citet{ivshina2022tess}. The transit midpoint reported by \citet{ivshina2022tess} occur approximately 2.68 minutes earlier than that from \citet{ehrenreich2020nightside}. Since the $K_{p}$ value for WASP-76b is $196.52 \pm 0.94 \; \mathrm{km\: s^{-1}}$ \citep{ehrenreich2020nightside} we can use Equation \ref{eq:simplified} to approximate the corresponding velocity offset. Using Equation \ref{eq:simplified}, a $2.68$-minute ephemeris offset is consistent with a $\sim 1.27\; \mathrm{km\: s^{-1}}$ velocity offset. Thus, our results agree with our analytical solution as presented in Equation \ref{eq:dvshift_final} and Figures \ref{fig:calcium} and \ref{fig:sodium}. 

\section{Discussion} \label{sec: discussion}

Various observational studies have documented Doppler blueshifts attributed to winds \citep[e.g.,][]{wyttenbach2015spectrally, casasayas2019atmospheric, hoeijmakers2019spectral, hoeijmakers2020high, bourrier2020hot, cabot2020detection, gibson2020detection, nugroho2020detection, stangret2021obliquity, tabernero2021espresso, borsa2021atmospheric, kesseli2021confirmation, rainer2021gaps, langeveld2022survey}. Typically, these studies have reported Doppler blueshift detections ranging from approximately $\sim2-8 \;\mathrm{km}\:\mathrm{s^{-1}}$.

However, virtually none of these papers take into account the potential issue of incorrect orbital parameters. For example, various orbital solutions for KELT-9b result in offsets ranging from $3$ to $11$ minutes \citep{gaudi2017giant, wong2020exploring}. With a $K_{p}$ of $\sim 250\;\mathrm{km}\: \mathrm{s}^{-1}$  \citep{borsa2022high} and a period of $\sim 1.5 \; \mathrm{days}$ \citep{gaudi2017giant}, Equation \ref{eq:simplified} yields a typical velocity offset between $2$-$8\;\mathrm{km\: s^{-1}}$. These velocity shifts, attributed to ephemeris offsets for KELT-9b, align with the findings of \citet{asnodkar2022variable}, who observed a $ 7.2\; \mathrm{km\: s^{-1}}$ velocity offset when using updated transit midpoint timings shifted up to $10 \; \mathrm{minutes}$.

Similarly, for KELT-20b, typical offsets in the transit midpoint values across different orbital solutions range from $1$ to $12 \; \mathrm{minutes}$ \citep{lund2017kelt, talens2018mascara, patel2022empirical, ivshina2022tess, kokori2023exoclock}. With a $K_{p}$ of $\sim 170\;\mathrm{km\: s^{-1}}$ \citep{asnodkar2022variable} and a period of $\sim 3.5$ days \citep{petz2024pepsi}, using Equation \ref{eq:simplified} once more, we estimate our typical velocity offset to be between $0$-$3\; \mathrm{km\: s^{-1}}$. \citet{asnodkar2022variable} reported that for KELT-20b, a 10-minute shift in transit midpoint corresponded to a velocity shift of $2.1\;\mathrm{km\: s^{-1}}$. Our findings are consistent with theirs, depending on the specific values of $K_{p}$ and the period used. 

For WASP-76b, updated transit midpoints from \citet{ivshina2022tess} would increase the blue-shifts that several studies have attributed to day-to-night winds \citep{deibert2021detection, seidel2021into, gandhi2022spatially}.

In the case of KELT-9b's orbit, \citet{asnodkar2022variable} has mentioned that the uncertainty in eccentricity measurement of the orbit could also affect the measurement of day-night winds. It is typical for exoplanet researchers to fix the eccentricity to zero. One study found a slight eccentric orbit for  KELT-9b \citep{pino2022gaps}, whereas originally the eccentricity was reported as $e < 0.007$ \citep[e.g.,][]{gaudi2017giant, wong2020exploring}. \citet{pino2022gaps} reported an eccentricity of $e = 0.016 \pm 0.003$ and an angle of periastron $\omega = 150^{\circ\,+13}_{-11}$. Looking at Equation \ref{eq: vshift_ecc}, we can estimate a resulting velocity offset of $3.5\;\mathrm{km\: s^{-1}}$. \citet{asnodkar2022variable} concluded 
that reasonably small eccentricities ($0.001 < e < 0.2$) did not affect the velocities in their model for KELT-9b. Even though small eccentricity offsets did not matter in the case of KELT-9b, they may have an effect on other exoplanet cases. 

Looking at the \dataset[NASA Exoplanet Archive]{\doi{10.26133/NEA1}} \footnote{\url{https://exoplanetarchive.ipac.caltech.edu/}}, for exoplanets with eccentricities that are assumed to be zero and have upper limits, we see upper limits ranging between $0.01$ and $0.10$, with a median and mean of $0.05$. Based on these upper limits and the angle of periastron ($\omega$), these uncertainties could result in velocity offsets of several kilometers per second as seen in Figure \ref{fig: angle periastron}. 

In the case of exoplanets with non-zero eccentricities, these eccentricities have uncertainties mostly between $\pm 0.001$ and $\pm 0.1$. There is a smaller concentration of uncertainties higher than $\pm 0.1$ and less than $\pm 0.5$. We are mainly interested in uncertainties that cause significant velocity shifts of several $\mathrm{km\: s^{-1}}$ \citep{snellen2010orbital, deibert2021detection, kesseli2021confirmation,langeveld2022survey}. Using Equation \ref{eq: vshift_ecc}, we find that for eccentricities ranging from $0.01$ to $0.1$, approximately $10 \%$ of uncertainties would cause shifts over $2\;\mathrm{km\: s^{-1}}$. Therefore, these uncertainties are significant and emphasize the importance of considering both the eccentricity itself and the uncertainty in its measurement.

Overall, these findings highlight the complexity in interpreting Doppler blueshifts caused by day–night winds. The observed range of blueshifts, approximately $2$–$15\;\mathrm{km\;s^{-1}}$, shows the significant impact that atmospheric winds can have on detected signals. For KELT-9b, KELT-20b, and WASP-76b, uncertainties in the transit midpoint produce velocity offsets that match the observed blueshifts, supporting our predictions. These offsets can not only mimic wind-driven signals but may also cause true wind velocities to be underestimated, as shown by \citet{asnodkar2022variable}.
 
Our findings emphasize the need to consider orbital parameters for an accurate interpretation of exoplanet data and for understanding their atmospheres. Our research specifically found that using recent transit time measurements and accounting for eccentricity---when both can significantly impact results---is crucial for accurate analysis of velocity offsets and potential day-night winds on an exoplanet. However, uncertainties in orbital measurements from transit observations with TESS or ground-based telescopes remain unavoidable. These uncertainties can be mitigated by obtaining simultaneous observations or by consistently using the most up-to-date transit times.

\section{Conclusion} \label{sec: conclusion}

Due to the high orbital speeds of planets relative to their stars, transmission spectroscopy allows us to study the atmosphere of a transiting planet. Our study shows that even small offsets in the transit midpoint or eccentricity can cause significant velocity shifts in our CCF matrix or transmission spectra. These shifts in our simulated data match closely with the predictions from our analytical or semi-analytical equations. Specifically, Equation \ref{eq:dvshift_final} explains the relationship between transit midpoint offset and velocity shifts, while Equation \ref{eq: vshift_ecc} describes the connection between eccentricity offset and velocity shifts.

Examining the impact of ephemeris offsets, as shown in Figure \ref{fig:lineplots}, reveals that small timing errors can cause velocity shifts of several kilometers per second, comparable to the typical range reported for day–night winds. The velocity offset changes proportionately depending on the value of $K_{p}$. Regarding eccentricity offset, as shown in Figure \ref{fig:ecc line plots}, even slight deviations in eccentricity can cause significant velocity shifts, depending on the angle of periastron ($\omega$).

Our findings suggest that the Doppler shifts associated with winds, often mentioned in exoplanet studies, are similar in magnitude to those caused by small orbital parameter offsets. This highlights the importance of considering these offsets when interpreting velocity shifts in exoplanetary systems, as reliable detections of atmospheric winds from transmission spectroscopy depend on accurate and well-constrained orbital parameters.

\begin{acknowledgments}
\section*{Acknowledgments}
Y.J.M acknowledges support from the Nexus Scholars program in the College of Arts and Sciences at Cornell University.

R.J. acknowledges support from a Rockefeller Foundation Bellagio Residency. 

J.D.T was supported for this work by the TESS Guest Investigator Program G06165 and by NASA through the NASA Hubble Fellowship grant \#HST-HF2-51495.001-A awarded by the Space Telescope Science Institute, which is operated by the Association of Universities for Research in Astronomy, Incorporated, under NASA contract NAS5-26555.

E.K.D. acknowledges the support of the Natural Sciences and Engineering Research Council of Canada (NSERC), funding reference number 568281-2022.

EdM acknowledges support from STFC award ST/X00094X/1.
Based on observations obtained at the international Gemini Observatory, a program of NSF NOIRLab, which is managed by the Association of Universities for Research in Astronomy (AURA) under a cooperative agreement with the U.S. National Science Foundation on behalf of the Gemini Observatory partnership: the U.S. National Science Foundation (United States), National Research Council (Canada), Agencia Nacional de Investigaci\'{o}n y Desarrollo (Chile), Ministerio de Ciencia, Tecnolog\'{i}a e Innovaci\'{o}n (Argentina), Minist\'{e}rio da Ci\^{e}ncia, Tecnologia, Inova\c{c}\~{o}es e Comunica\c{c}\~{o}es (Brazil), and Korea Astronomy and Space Science Institute (Republic of Korea).

Based on observations obtained through the Gemini Remote Access to CFHT ESPaDOnS Spectrograph (GRACES). ESPaDOnS is located at the Canada-France-Hawaii Telescope (CFHT), which is operated by the National Research Council of Canada, the Institut National des Sciences de l’Univers of the Centre National de la Recherche Scientifique of France, and the University of Hawai’i. ESPaDOnS is a collaborative project funded by France (CNRS, MENESR, OMP, LATT), Canada (NSERC), CFHT and ESA. ESPaDOnS was remotely controlled from the international Gemini Observatory, a program of NSF NOIRLab, which is managed by the Association of Universities for Research in Astronomy (AURA) under a cooperative agreement with the U.S. National Science Foundation on behalf of the Gemini partnership: the U.S. National Science Foundation (United States), the National Research Council (Canada), Agencia Nacional de Investigación y Desarrollo (Chile), Ministerio de Ciencia, Tecnología e Innovación (Argentina), Ministério da Ciência, Tecnologia, Inovações e Comunicações (Brazil), and Korea Astronomy and Space Science Institute (Republic of Korea).
\end{acknowledgments}

\facilities{Exoplanet Archive, Gemini:Gillett}
\software{astropy \citep{robitaille2013astropy, price2018astropy, The_Astropy_Collaboration_2022}, Matplotlib \citep{hunter2007matplotlib}, Numpy \citep{harris2020array}, OPERA \citep{martioli2012open}, pandas \citep{reback2020pandas}, SciPy \citep{virtanen2020scipy}.}

\bibliographystyle{aasjournal}
\bibliography{PaperDraft_FinalDraft}
\end{document}